\documentclass[12pt]{article}
\usepackage{graphicx}
\usepackage{amsmath}

\begin{document}

\title{Decay of charged fields in de Sitter spacetime}
\author{A. A. Smirnov\thanks{%
e-mail: saa@dfn.if.usp.br} \\
Institute of Physics of University of Sao Paulo}
\maketitle

\begin{abstract}
We study the decay of charged scalar and spinor fields around
Reissner-Nordstr\"{o}m black holes in de Sitter spacetime through
calculations of quasinormal frequencies of the fields. The influence of the
parameters of the black hole (charge, mass), of the decaying fields (charge,
spin), and of the spacetime (cosmological constant) on the decay is analyzed.

The analytic formula for calculation quasinormal frequencies for a large
multipole number (eikonal approximation) is derived both for the spinor and
scalar fields.
\end{abstract}

\section{Introduction}

It is well known that dynamical evolution of field perturbation on a black
hole background can be conventionally divided into three stages \cite{fn98}.
The first stage is an initial wave burst coming directly from the source of
perturbation and is dependent on the initial form of the original field
perturbation. The characteristic feature of the second stage is the damped
oscillations whose frequencies and damping times are defined by the
structure of the background spacetime and are independent of the initial
perturbation. This stage can be accurately described in terms of the
discrete set of quasinormal modes (QNM). And the third stage is an
asymptotic tail behavior of the waves at very late time which is caused by
backscattering of the gravitational field.

The original interest to the study of QNMs is arising from the possibility
to observe quasinormal ringing with the use of gravitational wave detectors
as it follows from theoretical predictions (for review see, e.g., \cite
{ks99-n99} and references therein). Recently the interest to the study of
QNMs has been reinforced in connection with their possible relation to the
thermodynamic properties of black holes in loop quantum gravity \cite
{h98,d03}.

A new application of the quasinormal mode spectrum has also arisen from
superstring theory \cite{m98,w98,rs99}. According to the AdS/CFT
correspondence, a large static black hole in asymptotically AdS spacetime
corresponds to a thermal state in CFT, and the decay of the test field in
the black hole spacetime corresponds to the decay of the perturbed state in
CFT, such that the quasinormal frequencies define the thermalization time
scale \cite{hh00}. Therefore, many authors focused their attention on the
studies of QNMs for different asymptotically AdS black holes \cite
{hh00,cm01,bss02,k02c,ckl03,k04}.

There exist observational evidences that the universe is described by the
general relativity equations with the positive cosmological constant \cite
{p97,cds98,g98}. That observation attracted considerable researches'
attention to the study of QNMs in asymptotically de Sitter spacetimes \cite
{of91,mn02,cl03,m03,b03,k03b,z04,kz04,j04}.

In \cite{of91,mn02} the calculation of the quasinormal frequencies for the
gravitational perturbations of the Schwarzschild de Sitter (SdS) black hole
was carried out. In \cite{k03b} the lower overtones of QNMs for higher
dimensional SdS black holes were calculated. In \cite{kz04} the total
spectrum of QNMs was obtained for the SdS black holes by numerical
calculations and the excellent coincidence with the 6th order WKB method was
shown. In \cite{z04} the low-laying quasinormal frequencies of the SdS black
hole for fields of different spin were calculated by the use the 6th order
WKB and the P\"{o}shl-Teller potential approximations. In \cite{cl03} the
authors derived an explicit expression for calculation of QNMs for the case
of a near extremal SdS black hole. In \cite{m03} that expression was
generalized to near extremal higher dimensional SdS and
Reissner-Nordstr\"{o}m de Sitter (RNdS) black holes. In \cite{b03} an
analytical method was developed to study the quasinormal mode spectrum of
SdS black holes for the scalar, electromagnetic, and gravitational fields in
the limit of nearly equal black hole and cosmological radii.

The QNMs of the massless uncharged Dirac fields for the RNdS black hole are
studied in \cite{j04} using the P\"{o}shl-Teller potential approximation. It
was found that the magnitude of the imaginary part of the quasinormal
frequencies decreases as the cosmological constant or the orbital angular
momentum increases, but it increases as the charge or the overtone number
increases. We note although that first the Dirac quasinormal modes were
evaluated in \cite{c03} for Schwarzschild black hole spacetimes.

Perturbations of a charged massless and massive scalar field was by
calculation of its QNMs in the Reissner-Nordstr\"{o}m (RN), were
investigated in \cite{k02} and \cite{k02b}, respectively. It was found that
the neutral perturbations dominate at the stage of the ``final ringdown''.
In \cite{zz04b,zz04} the spin $1/2$ Dirac particles with the positive,
negative, and zero charge $e$ in the presence of the RN\ black holes with
charge $Q$ was investigated. It was demonstrated that in late times the
neutral perturbations dominate when $eQ>0$ and the charged perturbations
dominate when $eQ<0$.

The decay of fields, which interact electromagnetically with the charge of a
black hole in de Sitter spacetime, was not considered till now. The
objective of the present paper is to study the decay of charged fields of
different spin in RNdS background and analyze how the decay is influenced by
a total set of parameters: the charge $Q$ of the black hole, the
cosmological constant $\Lambda $, and the charge $e$ of the decaying field.

The paper is organized as follows. In Sect. 2 we consider the spinor field
decay, in Sect. 3 we consider the scalar field decay, and in Sect. 4 we
summarize the results obtained.

\section{Decay of the spinor field}

In Schwarzschild coordinates, the metric for the Reissner-Nordstr\"{o}m de
Sitter black hole can be expressed as follows
\begin{equation}
ds^{2}=-fdt^{2}+f^{-1}dr^{2}+r^{2}\left( d\theta ^{2}+\sin ^{2}\theta
d\varphi ^{2}\right)  \label{eq201}
\end{equation}
\begin{equation}
f=1-\frac{2M}{r}+\frac{Q^{2}}{r^{2}}-\frac{\Lambda }{3}r^{2}  \label{eq202}
\end{equation}
where the parameters $M$, $Q$ are the mass and the charge of the black hole,
respectively, and $\Lambda $ is the cosmological constant. We introduce the
tortoise coordinate $r_{\ast }=\int f^{-1}dr$. Then we write the metric
function (\ref{eq202}) in the form
\begin{equation*}
f=\frac{\Lambda }{3r^{2}}\left( r-r_{-}\right) \left( r-r_{+}\right) \left(
r_{c}-r\right) \left( r-r_{b}\right) \,,
\end{equation*}
where $r_{c}$ is the cosmological horizon, $r_{-}$ is the inner event
horizon, and $r_{+}$ is the outer event horizon ($r_{b}$ can be found
through $r_{-}$, $r_{+}$, $r_{c}$). One can find that the tortoise
coordinate $r_{\ast }$ can be expressed as follows
\begin{eqnarray}
&&r_{\ast }=\frac{1}{2}\left[ \frac{1}{\kappa _{-}}\ln \left( r-r_{-}\right)
+\frac{1}{\kappa _{+}}\ln \left( r-r_{+}\right) \right.  \label{eq208} \\
&&\left. -\frac{1}{\kappa _{c}}\ln \left( r_{c}-r\right) +\frac{1}{\kappa
_{b}}\ln \left( r-r_{b}\right) \right] \,,  \notag \\
&&\kappa _{a}=\left. \frac{1}{2}\frac{df}{dr}\right| _{r=r_{a}},\;a=\left(
-,+,c,b\right) \,.  \notag
\end{eqnarray}
From (\ref{eq208}) one can see that $r_{\ast }\rightarrow \infty $ as $%
r\rightarrow r_{c}$\ and $r_{\ast }\rightarrow -\infty $\ as $r\rightarrow
r_{+}$.

The wave equation of the massless Dirac field can be written as
\begin{equation}
\left[ \gamma ^{a}e_{a}^{\;\mu }\left( \partial _{\mu }+\Gamma _{\mu
}+eA_{\mu }\right) \right] \Psi =0  \label{eq301}
\end{equation}
where $\gamma ^{a}$ are the Dirac matrices, $e_{a}^{\;\mu }$ is the inverse
of the tetrad $e_{\mu }^{\;a}$ , $\Gamma _{\mu }=\frac{1}{8}\left[ \gamma
^{a},\gamma ^{b}\right] e_{a}^{\;\nu }e_{b\nu ;\mu }$ is the spin
connection. We take the tetrad $e_{\mu }^{\;a}$ as \cite{bw57}
\begin{equation}
e_{\mu }^{\;a}=diag\left( f^{1/2},f^{-1/2},r,r\sin \theta \right) \,.
\label{eq302}
\end{equation}
We define $\Phi =f^{-1/4}\Psi $, then using the ansatz
\begin{equation}
\Phi =\left(
\begin{array}{c}
\frac{iG^{\left( \pm \right) }\left( r\right) }{r}\phi _{jm}^{\pm }\left(
\theta ,\varphi \right) \\
\frac{F^{\left( \pm \right) }\left( r\right) }{r}\phi _{jm}^{\mp }\left(
\theta ,\varphi \right)
\end{array}
\right) e^{-i\omega t}  \label{eq303}
\end{equation}
where
\begin{equation*}
\phi _{jm}^{+}=\left(
\begin{array}{c}
\sqrt{\frac{j+m}{2j}}Y_{l}^{m-1/2} \\
\sqrt{\frac{j-m}{2j}}Y_{l}^{m+1/2}
\end{array}
\right) \;\text{for }j=l+\frac{1}{2}
\end{equation*}
\begin{equation*}
\phi _{jm}^{-}=\left(
\begin{array}{c}
\sqrt{\frac{j+1-m}{2j+2}}Y_{l}^{m-1/2} \\
-\sqrt{\frac{j+1+m}{2j+2}}Y_{l}^{m+1/2}
\end{array}
\right) \;\text{for }j=l-\frac{1}{2}
\end{equation*}
we find for both signs of $F$ and $G$,
\begin{eqnarray}
\frac{d^{2}F}{dr_{\ast }^{2}}+\left( \omega ^{2}-V_{1}\right) F &=&0
\label{eq305} \\
\frac{d^{2}G}{dr_{\ast }^{2}}+\left( \omega ^{2}-V_{2}\right) G &=&0
\label{eq306}
\end{eqnarray}
where
\begin{eqnarray}
&&V_{1,2}=\pm \frac{dW}{dr_{\ast }}+W^{2},\;W=\frac{\left| k\right| \sqrt{f}%
}{r}-\frac{eQ}{r}\,,  \label{eq307} \\
&&k=j+\frac{1}{2},\;j=l+\frac{1}{2}\;\text{for }V_{1}\,,  \notag \\
&&k=-\left( j+\frac{1}{2}\right) ,\;j=l-\frac{1}{2}\;\text{for }V_{2}\,.
\notag
\end{eqnarray}
The quasinormal modes are defined as solutions of (\ref{eq305}), (\ref{eq306}%
) satisfying the boundary conditions
\begin{equation}
F\left( r_{\ast }\right) \sim A_{\pm }e^{\pm i\omega r_{\ast }},\;G\left(
r_{\ast }\right) \sim B_{\pm }e^{\pm i\omega r_{\ast }},\;r_{\ast
}\rightarrow \pm \infty \,,  \label{eq308}
\end{equation}
supposing \textrm{Re}$\omega >0$, that corresponds to purely in-going waves
at the event horizon and purely out-going waves the cosmological horizon.


\begin{figure}[!ht]
\centering
\includegraphics[width=3.5 in]{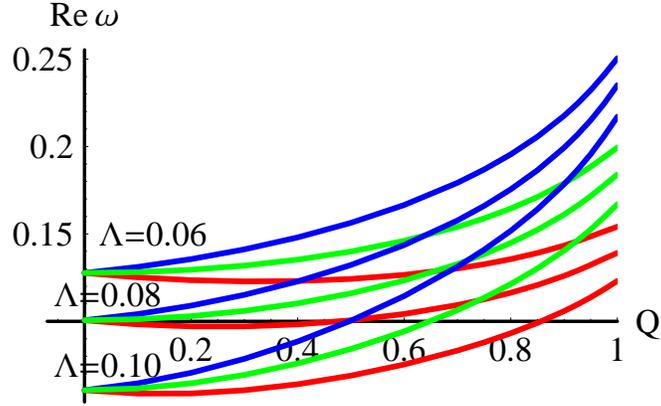}
\caption{QN frequencies of the spinor field; red: $e=0.1$, blue:
$e=-0.1$, green: $e=0$.}
\end{figure}

\begin{figure}[!ht]
\centering
\includegraphics[width=3.5 in]{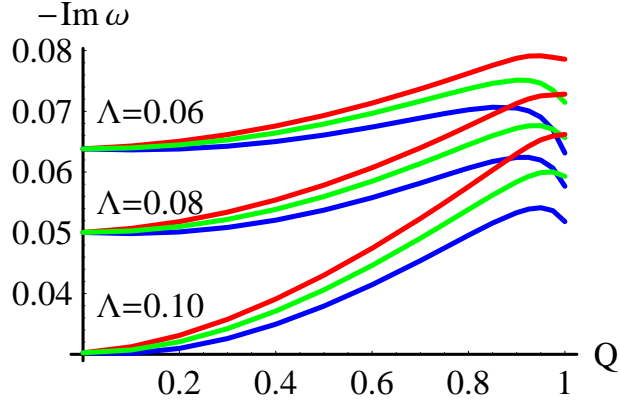}
\caption{Imaginary part of $\protect\omega $ of the spinor field;
red: $e=0.1$, blue: $e=-0.1$, green: $e=0$.}
\end{figure}

\begin{figure}[!ht]
\centering
\includegraphics[width=3.5 in]{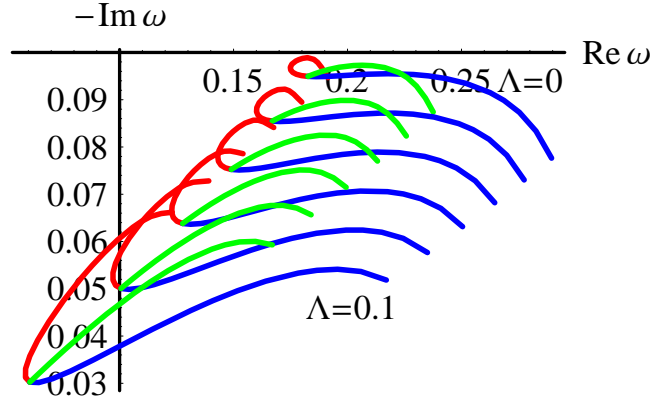}
\caption{QNM of the spinor field parametrized by $Q$ for $\Lambda =0$, $0.2$%
, $0.4$, $0.6$, $0.8$, $0.1$; red: $e=0.1$, blue: $e=-0.1$,
green: $e=0$.}
\end{figure}

\begin{figure}[!ht]
\centering
\includegraphics[width=3.5 in]{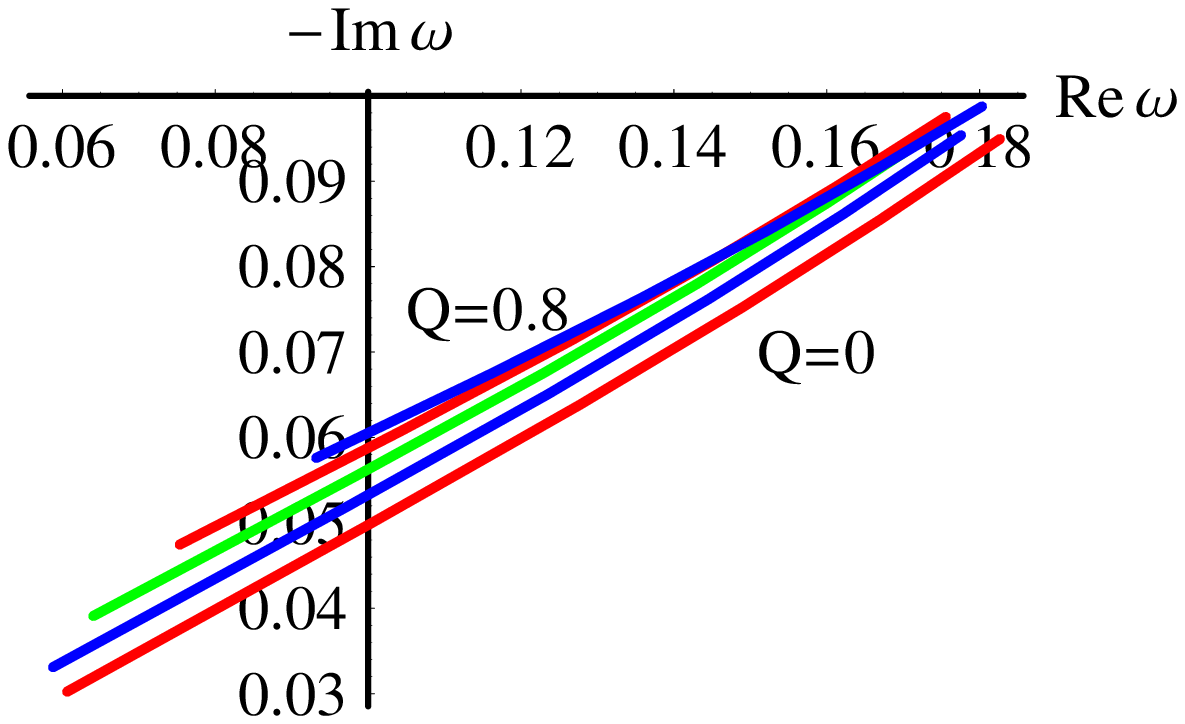}
\caption{QNM of the spinor field parametrized by $\Lambda $ for $e=0.1$, $%
Q=0 $, $0.2$, $0.4$, $0.6$, $0.8$.}
\end{figure}

\begin{figure}[!ht]
\centering
\includegraphics[width=3.5 in]{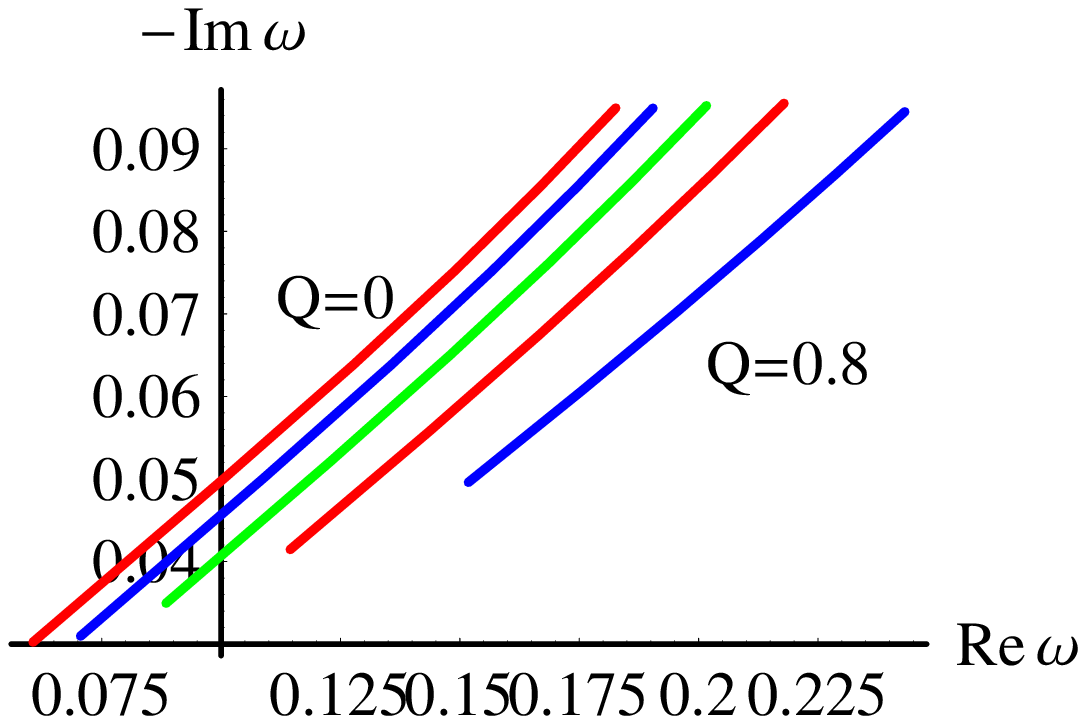}
\caption{QNM of the spinor field parametrized by $\Lambda $ for $e=-0.1$, $%
Q=0$, $0.2$, $0.4$, $0.6$, $0.8$.}
\end{figure}


We notice that the potentials $V_{1}$, $V_{2}$ give the same quasinormal
frequencies \cite{j04}, \cite{zz04}, therefore in what follows we focus the
attention on $V_{1}$. The effective potential $V_{1}$ (\ref{eq307}) is a
smooth function of $r_{\ast }$. It tends to constant values at the
cosmological horizon and at the outer event horizon and has a maximum near
the outer event horizon that allows to use the WKB method to calculate
quasinormal modes. To compute QNMs from Eqs. (\ref{eq305}), (\ref{eq307}) we
apply the WKB method of the sixth order \cite{k03}.

We present graphically the results for the fixed overtone number $n=0$ and $%
k=1$ at Figs. 1-5 for the charge of the spinor field $e=0$, $\pm
0.1$. It is shown the dependence of QN frequencies (Fig. 1) and
their damping rates (Fig. 2) on the charge of the black hole $Q$
for different values of the cosmological constant $\Lambda $. On
Figs. 1, 2 the dependences are shown for three values of the
charge of the spinor field $e=0$, $\pm 0.1$, which on the figures
looks like a triplet of curves beginning at the same point at
$Q=0$. The curves of different densities on the graphs correspond
to different $e$. On Fig. 1 the upper curve in the triplets corresponds to $%
e=-0.1$, the lower curve in the triplets corresponds to $e=0.1$,
the middle curve corresponds to $e=0$. On Fig. 2 the lower curve
in the triplets corresponds to $e=-0.1$, the upper curve in the triplets corresponds to $%
e=0.1$, the middle curve corresponds to $e=0$. The QN frequencies
are monotonically growing with $Q$ for $e=0$, $-0.1$, although for
$e=0.1$ the behavior of the curves is qualitatively different.
The curves for $e=0.1$ have a minimum at some point $Q_{0}$. The
value of $Q_{0}$\ is shifting to less values as $\Lambda $ grows.
Besides, Fig. 1 demonstrates that the QN frequencies become lower
as $\Lambda $ increasing. The position of a curve
in the triplet says that the QN frequencies is lower for the field charge $%
e>0$ and higher for $e<0$ ($Q>0$). The curves displaying QNM damping rate
are similar for all $\Lambda $ and $e$. They have a maximum at $Q_{\max
}\sim 0.9$. Until reaching $Q_{\max }$ the curves are slowly growing and
after reaching $Q_{\max }$ they are sharply falling down. The greater $%
\Lambda $ is the greater the rise of the curves is until reaching $Q_{\max }$%
. The value of $Q_{\max }$\ is shifting to greater values as $\Lambda $
increasing. The position of a curve in the triplet says that the decay of
the field is slower for the field charge $e<0$ and faster for $e>0$ ($Q>0$).
The decay of the spinor fields becomes slower as $\Lambda $ grows. Since the
charges $e$, $Q$ appear in $V_{1}$ only as $e^{2}$, $Q^{2}$ or$\ eQ$ one can
also conclude that for $Q<0$ the conclusions become upside-down. These
observations generalize results the work \cite{zz04} obtained for
asymptotically flat spacetimes. The peculiarities of the dependence of $%
\mathrm{Im}\omega $, \textrm{Re}$\omega $ on $Q$ are more obvious when
depicted on diagrams of the dependence of $\mathrm{Im}\omega $ on \textrm{Re}%
$\omega $ and parametrized by the black hole charge $Q$. These curious
pictures are presented on\ Fig. 3.

The dependence of \textrm{Im}$\omega $ on \textrm{Re}$\omega $ is
parametrized by the cosmological constant $\Lambda $, which is changing
along the curves, is depicted on Figs. 4, 5 for the fixed black hole charge $%
Q$. The curves are approximately straight lines. This observation
generalizes conclusions the research \cite{j04} obtained for the
uncharged spinor field. Both the QN frequencies and their damping
rates are decreasing with $\Lambda $, although the manner of the
decrease for \textrm{Re}$\omega $ and \textrm{Im}$\omega $ is
different.

Making use of the first order WKB method we also found the asymptotic
formula of large $k$ for calculation quasinormal modes $\omega $,
\begin{equation}
\omega =C_{0}k-i\left( n+1/2\right) C_{i}^{sp}+C_{r}^{sp}+O\left( 1/k\right)
\label{eq330}
\end{equation}
where
\begin{eqnarray}
&&C_{0}=\left( \frac{Q^{2}}{r_{0}^{4}}-\frac{2M}{r_{0}^{3}}+\frac{1}{%
r_{0}^{2}}-\frac{\Lambda }{3}\right) ^{1/2}\,,  \label{eq331} \\
&&C_{i}^{sp}=\frac{1}{\sqrt{3}r_{0}^{3}}\left[ -42Q^{4}+3Q^{2}r_{0}\left(
42M-15r_{0}+2\Lambda r_{0}^{3}\right) +\right.  \notag \\
&&\left. r_{0}^{2}\left( 90M^{2}-60Mr_{0}+9r_{0}^{2}+6M\Lambda
r_{0}^{3}-\Lambda r_{0}^{4}\right) \right] ^{1/2}\,,  \notag \\
&&C_{r}^{sp}=-\frac{1}{2r_{0}^{2}}\left[ 4r_{1}^{sp}r_{0}C_{0}^{1/2}+\left(
r_{0}-3M\right) +\right.  \notag \\
&&\left. \frac{2}{3r_{0}^{2}}C_{0}^{-1/2}\left(
Q^{2}+eQr_{0}^{2}+r_{1}^{sp}\left( 3M-3r_{0}+2r_{0}^{3}\Lambda \right)
\right) \right] \,,  \notag
\end{eqnarray}
\begin{eqnarray}
&&r_{0}=\frac{3}{2}M\left( 1+\sqrt{1-\frac{8Q^{2}}{9M^{2}}}\right) \,,
\label{eq332} \\
&&r_{1}^{sp}=-\frac{1}{2}r_{0}^{2}C_{0}^{1/2}  \notag \\
&&\left( -30Q^{4}+90MQ^{2}r_{0}-63M^{2}r_{0}^{2}-33Q^{2}r_{0}^{2}\right.
\notag \\
&&-18eQ^{3}r_{0}^{2}+42Mr_{0}^{3}+30eMQr_{0}^{3}-6r_{0}^{4}  \notag \\
&&\left. +12eQr_{0}^{4}+6Q^{2}r_{0}^{4}\Lambda -6Mr_{0}^{5}\Lambda
+r_{0}^{6}\Lambda +2eQr_{0}^{6}\Lambda \right)  \notag \\
&&\left( -42Q^{4}+126MQ^{2}r_{0}-90M^{2}r_{0}^{2}-45Q^{2}r_{0}^{2}\right.
\notag \\
&&\left. +60Mr_{0}^{3}-9r_{0}^{4}+6Q^{2}r_{0}^{4}\Lambda -6Mr_{0}^{5}\Lambda
+r_{0}^{6}\Lambda \right) ^{-1}\,.  \notag
\end{eqnarray}
By setting $Q=0$ one can make sure that the formula (\ref{eq330}) reproduces
the corresponding formula of the paper \cite{z04} obtained for the SdS black
hole. We note that \textrm{Im}$\omega $ does not depend on the charge of the
spinor field $e$ in the limit of large $k$. Therefore the damping rate is
equal for the arbitrary spinor field charge for large enough multipole
moment. The dependence of \textrm{Re}$\omega $ on $k$ is linear for large $k$%
. The slope of the line on the plane $\left( \mathrm{Re}\omega ,k\right) $
is defined by the coefficient $C_{0}$. These conclusions are confirmed by
the results of \cite{zz04} where the particular case $\Lambda =0$ was
considered.

\begin{figure}[!ht]
\centering
\includegraphics[width=3.5 in]{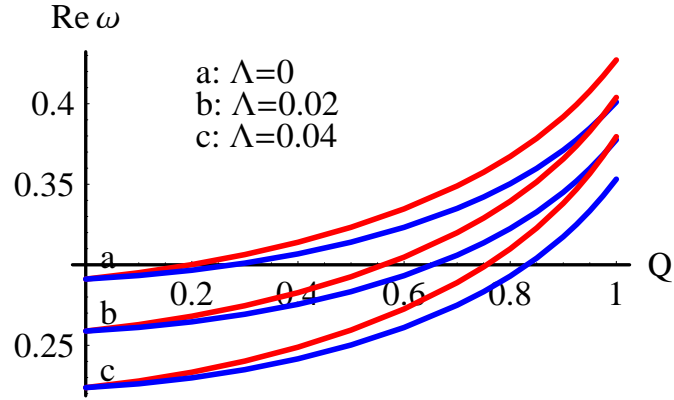}
\caption{QN frequencies of the scalar field; red: $e=0.1$, blue:
$e=0.05$.}
\end{figure}

\begin{figure}[!ht]
\centering
\includegraphics[width=3.5 in]{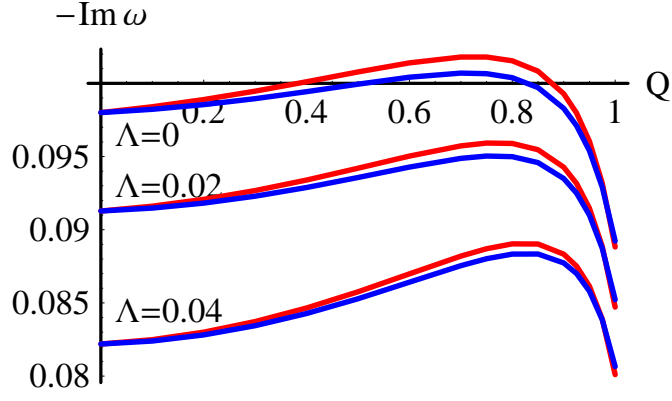}
\caption{Imaginary part of $\protect\omega $ of the scalar field; red: $%
e=0.1 $, blue: $e=0.05$.}
\end{figure}

\begin{figure}[!ht]
\centering
\includegraphics[width=3.5 in]{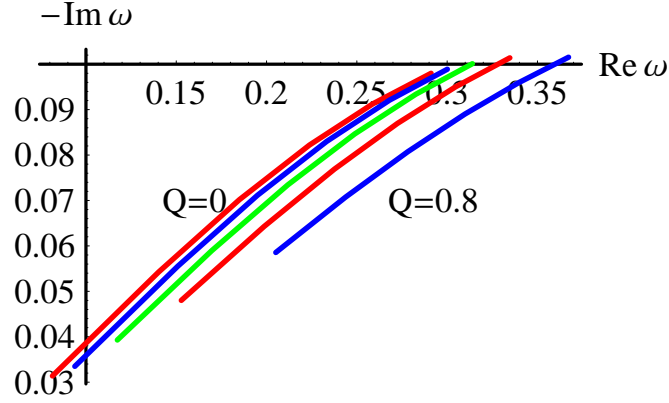}
\caption{QNM of the scalar field parametrized by $\Lambda $ for $Q=0$, $0.2$%
, $0.4 $, $0.6$, $0.8$; $e=0.1$.}
\end{figure}

\section{Decay of the scalar field}

The wave equation of the complex scalar field has the form \cite{he73},
\begin{equation}
\phi _{;ab}-ieA_{a}g^{ab}\left( 2\phi _{;b}-ieA_{b}\phi \right)
-ieA_{a;b}g^{ab}=0  \label{eq203}
\end{equation}
Here we choose the electromagnetic potential as $A_{t}=-Q/r$. Then,
decomposing the wave function $\phi $ into spherical harmonics $\phi
=\sum_{l,m}u^{l}{}_{m}\left( t,r\right) Y^{l}{}_{m}\left( \theta ,\varphi
\right) /r$, the wave equation for each multipole moment takes the form
\begin{equation}
\partial _{t}^{2}u+2ie\frac{Q}{r}\partial _{t}u-\partial _{r_{\ast
}}^{2}u+Vu=0  \label{eq204}
\end{equation}
\begin{equation}
V=f\left( \frac{l\left( l+1\right) }{r}+\frac{2M}{r^{3}}-\frac{2Q^{2}}{r^{4}}%
-\frac{2\Lambda }{3}\right) -\frac{e^{2}Q^{2}}{r^{2}}  \label{eq205}
\end{equation}
where $r_{\ast }$ is the tortoise coordinate and $u\left( t,r\right)
=e^{-i\omega t}u\left( r\right) $. Therefore we come to the equation for $%
u\left( r\right) $,
\begin{eqnarray}
&&\partial _{r_{\ast }}^{2}u+\left[ \omega ^{2}-V_{eff}\right] u=0\,,
\label{eq206} \\
&&V_{eff}=V+\frac{2eQ\omega }{r}\,.  \notag
\end{eqnarray}
Supposing \textrm{Re}$\omega >0$ solutions of (\ref{eq206}) have to satisfy
the boundary conditions
\begin{equation}
u\left( r_{\ast }\right) \sim C_{\pm }e^{\pm i\omega r_{\ast }},\;r_{\ast
}\rightarrow \pm \infty \,.  \label{eq206a}
\end{equation}

To calculate quasinormal modes we apply the WKB method of the third order
\cite{iw87} in the scalar field case. The motivation about the use of WKB
method is the same as that in Sect. 2. The use of the calculation method of
higher order \cite{k03} is overly cumbersome as in the case of scalar field $%
V_{eff}$ includes $\omega $ so that it is necessary to solve simultaneously
the equation for the search of an extremum and the equations of the WKB
method itself.

We present graphically the results for the fixed overtone number $n=0$ and
multipole number $l=1$ on Figs. 6-8. It is shown the dependence of QN
frequencies (Fig. 6) and their damping rates (Fig. 7) on the charge of the
black hole $Q$ for different values of the cosmological constant $\Lambda $.
On Figs. 6, 7 the dependences are shown for two values of the charge of the
scalar field $e=0.1$, $0.05$, which on the figures looks like a couple of
curves beginning at the same point at $Q=0$. The upper curve in the couples
corresponds to $e=0.1$, the lower curve lower one of the couples corresponds
to $e=0.05$.

The QN frequencies monotonically grow with $Q$ and become lower as $\Lambda $
increasing. The QN frequencies are higher for the fields with greater charge
$e$. The curves displaying the QNM damping rates have a pronounced maximum
which is located at $Q_{\max }\simeq 0.8$ and a location of the maximum is
shifting from $Q\simeq 0.7$.to $Q\simeq 0.9$ as $\Lambda $ increasing. Until
reaching $Q_{\max }$ the curves are slowly growing and after reaching $%
Q_{\max }$ they are sharply flowing down. Fig. 7 shows that the field decays
faster if it has greater charge. And similar to the spinor case the decay of
the scalar fields becomes slower as $\Lambda $ grows. Both the QN
frequencies and the damping rates are decreasing with $\Lambda $, although
the manner of the decrease is different.

Fig. 8 displays QNMs for different fixed $Q$. The curves on Fig. 8 are
parametrized by the cosmological constant which is changing along the
curves. We note that \textrm{Im}$\omega $ is related to \textrm{Re}$\omega $
not linearly in contrast to the spinor field case (Figs. 4, 5).

Making use of the first order WKB method it is also possible to find the
asymptotic formula of large $l$ for calculation quasinormal modes $\omega $
for the scalar field,
\begin{equation}
\omega =C_{0}\left( l+1/2\right) -i\left( n+1/2\right)
C_{i}^{sc}+C_{r}^{sc}+O\left( 1/l\right)  \label{eq210}
\end{equation}
where
\begin{eqnarray}
&&C_{i}^{sc}=\frac{1}{3}r_{0}^{-3}\left\{ -126Q^{4}r_{0}^{2}+Q^{2}r_{0}^{2}%
\left[ 378Mr_{0}+r_{0}^{2}\left( -135+18\Lambda r_{0}^{2}\right) \right]
+\right.  \label{eq211} \\
&&\left. r_{0}^{3}\left[ -270M^{2}r_{0}-6Mr_{0}^{2}\left( -30+3\Lambda
r_{0}^{2}\right) +r_{0}^{3}\left( -27+3\Lambda r_{0}^{2}\right) \right]
\right\} ^{1/2}\,\,,  \notag \\
&&C_{r}^{sc}=\frac{r_{1}^{sc}}{r_{0}^{3}}\left( -1-\frac{2Q^{2}}{r_{0}^{2}}+%
\frac{3M}{r_{0}}\right) C_{0}^{-1}\,,  \notag
\end{eqnarray}
\begin{eqnarray}
&&r_{1}^{sc}=-r_{0}\left(
-6Q^{4}+21MQ^{2}r_{0}-18M^{2}r_{0}^{2}-9Q^{2}r_{0}^{2}+\right.  \label{eq212}
\\
&&\left. 15Mr_{0}^{3}-3r_{0}^{4}+2Q^{2}\Lambda r_{0}^{4}-3M\Lambda
r_{0}^{5}+\Lambda r_{0}^{6}\right) \times  \notag \\
&&\left( 42Q^{4}-12MQ^{2}r_{0}+90M^{2}r_{0}^{2}+45Q^{2}r_{0}^{2}\right.
\notag \\
&&\left. -60Mr_{0}^{3}+9r_{0}^{4}-6Q^{2}\Lambda r_{0}^{4}+6M\Lambda
r_{0}^{5}-\Lambda r_{0}^{6}\right) \,.  \notag
\end{eqnarray}
In expressions (\ref{eq210})-(\ref{eq212}) $C_{0}$, $r_{0}$ are
the same as those in formulas (\ref{eq331}), (\ref{eq332}) of
Sect. 2. By setting $Q=0$ one can make sure that the formula
(\ref{eq330}) reproduces the corresponding formula of the paper
\cite{z04} obtained for the SdS black hole for the scalar field.
We note that the asymptotic formula for the RN black hole in
asyptotically flat spacetime for the charged scalar field was not
found till now. The formlula for such a particular case can be
obtained by setting $\Lambda=0$ in  (\ref{eq210})-(\ref{eq212}).

\section{Conclusions}

We considered the decay of the charged spinor and scalar field near the
Reissner-Nordstr\"{o}m black hole in de Sitter spacetime. We calculated the
quasinormal frequencies of the fields and their damping rates for the lower
overtone number. The analysis of the results allows to formulate the
following conclusions. a) In the spinor case for a fixed cosmological
constant $\Lambda $ the quasinormal frequencies of the charged field are
lower and the damping rates are higher than those of the neutral field if $%
eQ>0$ and vice versa if $eQ<0$; b) In the scalar case for a fixed
cosmological constant $\Lambda $ the fields possessing greater charge have
higher quasinormal frequencies and decay faster; c) The quasinormal
frequencies of both the spinor and scalar fields are lower in the spacetimes
with greater cosmological constant $\Lambda $. The decay of both the spinor
and scalar fields is slower in the spacetimes with greater cosmological
constant $\Lambda $; d) In the case of the charged spinor field the
dependence of \textrm{Im}$\omega $ on \textrm{Re}$\omega $ is approximately
linear on the diagram parametrized by $\Lambda $; e) In the case of the
charged scalar field the dependence of \textrm{Im}$\omega $ on \textrm{Re}$%
\omega $ is not linear on the diagram parametrized by $\Lambda $; f) The
dependence of \textrm{Im}$\omega $ and \textrm{Re}$\omega $ of the charged
spinor field on the black hole charge $Q$ is different for $e>0$ and $e<0$
for all values of $\Lambda $.

We derived the analytic formula for calculation of $\omega $ for large
values of the multipole number both for the spinor and scalar fields. The
formula for the spinor field shows that for asymptotically large multipole
number the damping rate does not depend on the charge $e$ of the spinor
field.

\end{document}